\documentclass{iaus}
\usepackage{graphicx,natbib,bm}

\newcommand{\EQ}{\begin{equation}}
\newcommand{\EN}{\end{equation}}
\newcommand{\EQA}{\begin{eqnarray}}
\newcommand{\ENA}{\end{eqnarray}}

\newcommand{\Sec}[1]{Section~\ref{#1}}

\newcommand{\Fig}[1]{Figure~\ref{#1}}

\newcommand{\bra}[1]{\langle #1\rangle}

\newcommand{\meanrho}{\overline{\rho}}

{}
{}
{}

{}
{}
\newcommand{\meanEMF}{\overline{\mbox{\boldmath ${\cal E}$}}{}}{}
{}
{}
{}
{}
{}
{}
\newcommand{\meanBB}{\overline{\mbox{\boldmath $B$}}{}}{}
{}
{}
{}
{}
{}
{}
{}
{}
\newcommand{\meanJJ}{\overline{\mbox{\boldmath $J$}}{}}{}
{}
\newcommand{\meanUU}{\overline{\bm{U}}}

{}
{}
{}

\newcommand{\meanB}{\overline{B}}

\newcommand{\meanU}{\overline{U}}

\newcommand{\meanJ}{\overline{J}}

{}

{}
{}

\newcommand{\yy}{\mbox{\boldmath $y$} {}}

\newcommand{\xx}{\bm{x}}

\newcommand{\uu}{\mbox{\boldmath $u$} {}}
\newcommand{\UU}{\mbox{\boldmath $U$} {}}

\def\bb{\bm{b}}
\newcommand{\BB}{\mbox{\boldmath $B$} {}}

\newcommand{\JJ}{\mbox{\boldmath $J$} {}}

\newcommand{\AAA}{\mbox{\boldmath $A$} {}}

\newcommand{\ff}{\mbox{\boldmath $f$} {}}

\newcommand{\grav}{\mbox{\boldmath $g$} {}}
\newcommand{\nab}{\mbox{\boldmath $\nabla$} {}}

\newcommand{\oo}{\mbox{\boldmath $\omega$} {}}

\newcommand{\SSSS}{\mbox{\boldmath ${\sf S}$} {}}

\newcommand{\DD}{{\rm D} {}}
\newcommand{\DDD}{{\cal D} {}}
\newcommand{\dd}{{\rm d} {}}

\def\Sh{\mbox{\rm Sh}}

\def\Pm{\mbox{\rm Pr}_M}
\def\Rm{\mbox{\rm Re}_M}

\def\cs{c_{\rm s}}

\def\qp{q_{\rm p}}

\def\qs{q_{\rm s}}

\def\kf{k_{\rm f}}
\def\epsf{\epsilon_{\rm f}}

\def\urms{u_{\rm rms}}

\def\nut{\nu_{\rm t}}

\def\etat{\eta_{\it t}}

\def\etaT{\eta_{\rm T}}

\def\Beq{B_{\rm eq}}

\def\half{{\textstyle{1\over2}}}

\def\onethird{{\textstyle{1\over3}}}

\newcommand{\G}{\,{\rm G}}

\newcommand{\kG}{\,{\rm kG}}

\newcommand{\s}{\,{\rm s}}

\newcommand{\cm}{\,{\rm cm}}

\newcommand{\Mm}{\,{\rm Mm}}

\newcommand{\yapj}[3]{ #1, {ApJ,} {#2}, #3}

\newcommand{\yapjl}[3]{ #1, {ApJ,} {#2}, #3}

\newcommand{\yan}[3]{ #1, {Astron.\ Nachr.,} {#2}, #3}

\newcommand{\yana}[3]{ #1, {A\&A,} {#2}, #3}

\newcommand{\ygafd}[3]{ #1, {Geophys.\ Astrophys.\ Fluid Dyn.,} {#2}, #3}

\newcommand{\yjetp}[3]{ #1, {Sov.\ Phys.\ JETP,} {#2}, #3}

\newcommand{\ymn}[3]{ #1, {MNRAS,} {#2}, #3}
\newcommand{\ynat}[3]{ #1, {Nature,} {#2}, #3}

\newcommand{\ysci}[3]{ #1, {Science,} {#2}, #3}
\newcommand{\ysph}[3]{ #1, {Solar Phys.,} {#2}, #3}

\newcommand{\ypre}[3]{ #1, {Phys.\ Rev.\ E,} {#2}, #3}

\newcommand{\yjour}[4]{ #1, {#2}, {#3}, #4}

\newcommand{\sapj}[1]{ #1, {ApJ}, submitted}
\newcommand{\san}[1]{ #1, {Astron.\ Nachr.}, submitted}
\newcommand{\sana}[1]{ #1, {A\&A}, submitted}
\newcommand{\smn}[1]{ #1, {MNRAS}, submitted}
\graphicspath{{./fig/}{./png/}}

\title[]
{Cycles and cycle modulations}

\author[Brandenburg \& Guerrero]
{Axel Brandenburg$^{1,2}$ \and Gustavo Guerrero$^{1,3}$}

\affiliation{
$^1$Nordita, Roslagstullsbacken 23,
SE-10691 Stockholm, Sweden, email: {\tt brandenb@nordita.org}
\\[\affilskip]
$^2$Department of Astronomy, Stockholm University,
SE-10691 Stockholm, Sweden
\\[\affilskip]
$^3$Solar Physics, HEPL, Stanford University, Stanford, CA 94305-4085, USA
}

\pubyear{2012}
\volume{286}
\pagerange{1--12}
\jname{Comparative magnetic minima}
\editors{C. H. Mandrini, eds.}
\begin{document}

\maketitle

\begin{abstract}
Some selected concepts for the solar activity cycle are briefly reviewed.
Cycle modulations through a stochastic $\alpha$ effect are being
identified with limited scale separation ratios.
Three-dimensional turbulence simulations with helicity and shear are
compared at two different scale separation ratios.
In both cases the level of fluctuations shows relatively little variation
with the dynamo cycle.
Prospects for a shallow origin of sunspots are discussed in terms of the
negative effective magnetic pressure instability.
Tilt angles of bipolar active regions are discussed as a consequence of
shear rather than the Coriolis force.
\end{abstract}

\firstsection

\section{Solar cycle}

The solar cycle manifests itself through spots at the Sun's surface.
To understand activity variations, we have to understand not only
their source, but also the detailed connection
between variations in the strength of the dynamo and its effect
on the number and size of sunspots.
In this paper, we address both aspects.

The physics of the solar cycle is not entirely clear.
The models that work best are not necessarily those that would emerge
from first principles.
Even the reason for the equatorward migration of the activity belts
is not completely clear.
Following Parker's seminal paper of 1955, this migration seemed to be 
a simple property of an $\alpha\Omega$ dynamo, i.e., a dynamo that works with
$\alpha$ effect and shear.
What matters for equatorward migration is not the $\Omega$ gradient in the
latitudinal direction, but that in the radial one, $\partial\Omega/\partial r$.
However, in the bulk of the convection zone, $\partial\Omega/\partial r$
is mostly positive.
This, together with an $\alpha$ effect of positive sign in the northern
hemisphere results in poleward migration \citep{Yos75}, which is not what
is observed.
On the other hand, according to the flux transport dynamos, magnetic fields
are advected by the meridional circulation.
Assuming that there is a coherent circulation with equatorward migration
at the bottom of the convection zone, this would then turn the dynamo wave
around so as to explain the solar butterfly diagram and that sunspots emerge
from progressively lower latitudes \citep{CSD95,DC99,GdG08}.
This requires that most of the field resides at the bottom of the
convection zone.
Moreover, the $\alpha$ effect is taken to be non-vanishing only near
the very top of the convection zone, i.e., the mean electromotive force
has to be written formally as a convolution of the mean magnetic field
with an integral kernel to account for this non-locality
\citep[see, e.g.,][]{BK07}.
Furthermore, from the observed tilt angles of bipolar regions
it is inferred that the Sun's magnetic field at the bottom
of the convection zone reaches strengths of the order of $100\kG$
\citep{DSC93}, which is nearly 100 times over the equipartition value.
Finally, there are assumptions about the turbulent magnetic diffusivity.
In all cases, the magnetic diffusivity in the evolution equation for the
toroidal field in the bulk of the convection zone is rather small,
below $10^{11}\cm^2\s^{-1}$ \citep[see, e.g.,][]{CC06}.
The magnetic diffusivity for the poloidal field is assumed to be 
larger and similar to the values expected from  mixing length 
theory (see below).

In any case, these assumptions are hardly in agreement with standard formulae
that the magnetic diffusivity is given by $\onethird\tau\urms^2$, where
$\tau$ is the turnover time and $\urms$ is the rms value of the turbulent
velocity.
The turnover time is $\tau=(\urms\kf)^{-1}$, where $\kf$ is the wavenumber
of the energy-carrying eddies.
This result for $\etat$ is well confirmed by simulations \citep{SBS08}.
For the Sun, mixing length theory appears to be reasonably good and gives
$\etat\approx(1$...$3)\times10^{12}\cm^2\s^{-1}$.
Also, in contrast to the assumptions of some flux transport dynamo models,
a strong degree of anisotropy of the $\eta$ tensor is not expected
from theory \citep{BRK11}.

An alternate approach is to use turbulent transport coefficients from theory,
which give rise to what is called a distributed dynamo, i.e., the induction
effects are non-vanishing and distributed over the entire convection zone.
In addition, there is the hypothesis that the near-surface shear layer may
be important for the equatorward migration \citep{B05}, but this has never
been confirmed by simulations either.
In any case, based on such models one would not expect there to be
a $100\kG$ magnetic field, but only a much weaker field of around
$0.3$--$1\kG$.
This calls then for an alternative explanation for the magnetic field
concentrations of up to $3\kG$ seen in sunspots and active regions.
Various proposals were already discussed in \cite{B05}, and meanwhile
there are direct numerical simulations (DNS) confirming the validity
of the physics assumed in one of those proposals.
This will be addressed in \Sec{sunspots}.

\section{Cycle modulation}

Early ideas for cycle modulations go back to \cite{Tav78}
who argued that the solar cycle may be a chaotic attractor.
This explanation became very popular in the following years
\citep{Ruz81,WCJ84}.
These ideas were elaborated upon in the framework of low-order
truncations of mean-field dynamo models, having in mind that
the same idea applies also to the underlying fully nonlinear
three-dimensional equations of magnetohydrodynamics.
Another line of thinking is that in mean-field theory (MFT) the physics of
the cycle models can be explained by random fluctuations in the turbulent
transport coefficients \citep{Cho92,Moss92,SSFM96,BS08}.
For high-dimensional attractors there is hardly any difference between
both approaches.
A completely different proposal for cycle modulation is related to variations
in the meridional circulation \citep{NMM11}.
This proposal still lacks verification from DNS of a dynamo whose cycle
period is indeed controlled by meridional circulation.
By contrast, fluctuations in the turbulent transport coefficients have
indeed been borne out by simulations \citep[see][]{BRRK08}.

To illustrate this, let us now consider a physical realization of a simple
$\alpha\Omega$ dynamo in a periodic domain.
In the language of MFT, this corresponds to solving the following set
of mean-field equations,
\EQ
{\partial\meanBB\over\partial t}=\nab\times\left(
\meanUU\times\meanBB+\meanEMF-\eta\mu_0\meanJJ\right)
\EN
in a Cartesian domain, $(x,y,z)$, in one dimension, $-\pi<z<\pi$, where
$\meanUU=\meanUU_S\equiv(0,Sx,0)$ is a linear shear flow velocity,
$\meanJJ=\nab\times\meanBB/\mu_0$ is the mean current density,
$\mu_0$ is the vacuum permeability, $\eta$ is the microphysical
(molecular) magnetic diffusivity, and
\EQ
\meanEMF=\alpha\meanBB-\etat\mu_0\meanJJ
\EN
is the mean electromotive force.
In DNS, on the other hand, one often solves the equations for an
isothermal gas with constant sound speed $\cs$,
\EQ
{\partial\BB\over\partial t}=\nab\times\left(
\UU\times\BB-\eta\mu_0\JJ\right),
\EN
together with corresponding equations governing the evolution of
the turbulent velocity $\UU$.

In the following we present results of simulations using shearing--periodic
boundary conditions.
To maintain the solenoidality of the magnetic field, we write
$\BB=\nab\times\AAA$ and solve for the magnetic vector potential $\AAA$.
Using in the following the velocity for the deviations from the shear flow,
$\UU$, our equations are
\EQ
\frac{\partial\AAA}{\partial t}+\UU_S\cdot\nab\AAA=-SA_y\xx+
\UU\times\BB+\eta \nab^2 \AAA,
\label{dAdt}
\EN
\EQ
\frac{\DDD\UU}{\DDD t} = -SU_x\yy -\cs^{2}\nab\ln{\rho} +\ff
+{1\over\rho}\left(\JJ\times\BB+\nab\cdot2\nu\rho\SSSS\right),
\label{dUdt}
\EN
\EQ
\frac{\DDD\ln\rho}{\DDD t}=  -\nab\cdot\UU,
\label{dRdt}
\EN
where $\DD/\DD t=\partial/\partial t+(\UU+\UU_S)\cdot\nab$ is the advective
derivative with respect to the total flow, $\UU+\UU_S$,
$\rho$ is the gas density, $\nu$ is the viscosity,
${\sf S}_{ij}=\half(U_{i,j}+U_{j,i})-\onethird\delta_{ij}\nab\cdot\UU$
is the trace-less rate of strain matrix, and $\ff$ is a forcing function
that drives both turbulence and a linear shear flow.
Alternatively, turbulence can also be the result of some instability
(Rayleigh-B\'enard instability, magneto-rotational instability, etc).
In the following we restrict ourselves to a random forcing function
with wavevectors whose modulus is in a narrow interval around an average
wavenumber $\kf$.
This has the additional advantage that we can arrange the forcing
function such that it has a part that is fully helical, i.e.,
$\nab\times\ff=\kf\ff$ (the part driving the shear flow is of course
non-helical).
Because of the presence of helicity, we should expect there to be an
$\alpha$ effect operating in the system, but if the number of turbulent
eddies in the domain is not very large, there can be significant fluctuations
in the resulting $\alpha$ effect.

\begin{figure}[t!]\begin{center}
\includegraphics[width=\columnwidth]{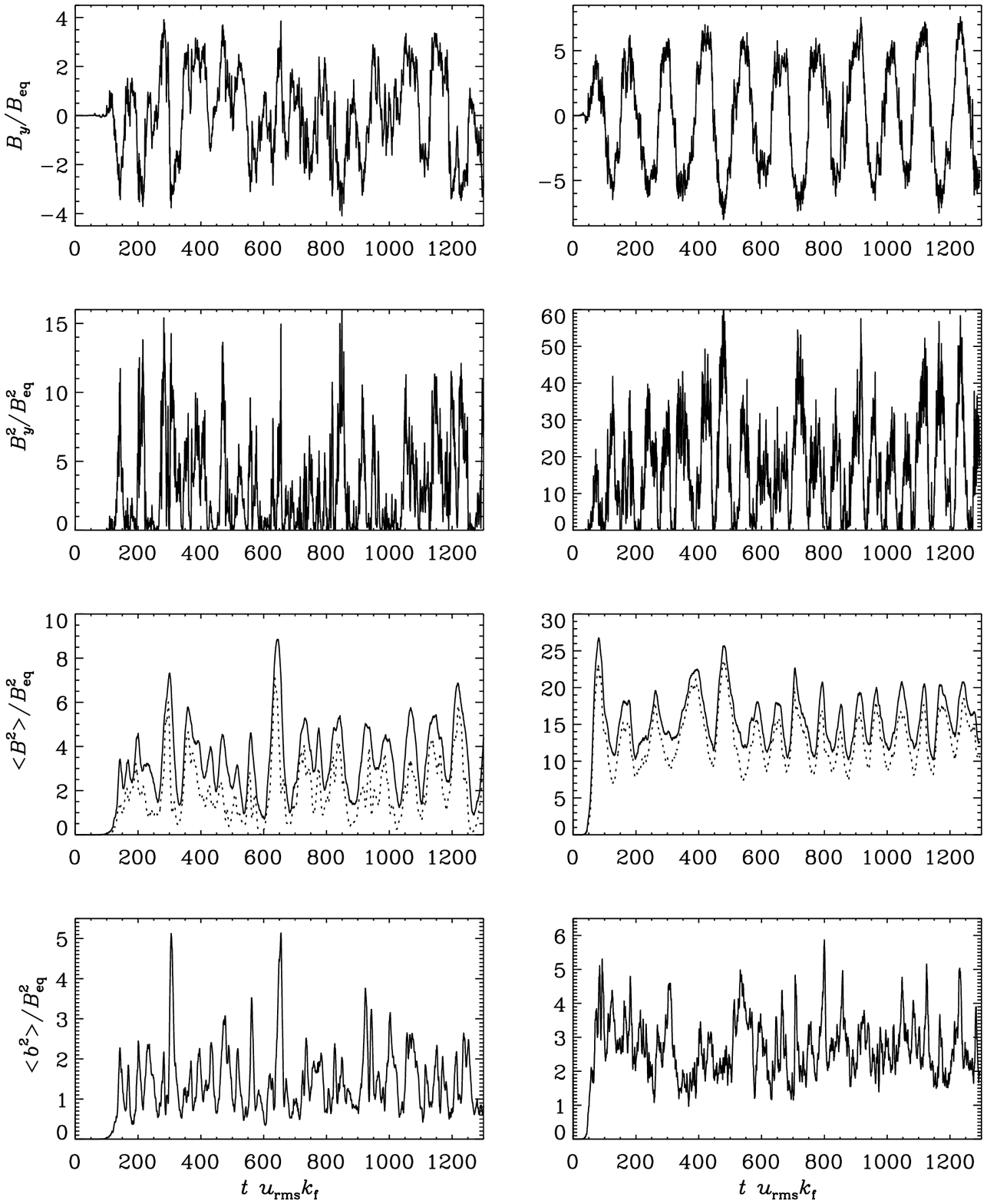}
\end{center}\caption[]{
Time sequences of $B_y$, $B_y^2$, $\bra{\BB^2}$ (solid line)
together with $\bra{\meanBB^2}$ (dotted line), and $\bra{\bb^2}$.
The field is always normalized by $\Beq$.
Here, $\Rm=22$ for $\kf/k_1=1.5$ (left column) and 
$\Rm=9$ for $\kf/k_1=2.2$ (right column).
In both cases, $\Pm=5$.
The shear parameter is $\Sh\approx-2$ in both cases.
}\label{pncomp2}\end{figure}

Important control parameters are the magnetic Reynolds and Prandtl numbers,
$\Rm=u_{\rm rms}/\eta\kf$ and $\Pm=\nu/\eta$.
In addition, there is the shear parameter defined here as $\Sh=S/\urms\kf$.
The smallest possible wavenumber in a triply-periodic domain of size
$L\times L\times L$ is $k_1=2\pi/L$.
For the purpose of presenting exploratory results, we restrict
ourselves here to a resolution of $64^3$ meshpoints.
We use the fully compressible {\sc Pencil Code} \footnote{
http://www.pencil-code.googlecode.com} for all our calculations.

In \Fig{pncomp2} we present the results of two simulations with
scale separation ratios $\kf/k_1$ of 1.5 and 2.2.
In both cases, $\kf/k_1$ is still relatively small,
but the difference in the results is already quite dramatic.
For $\kf/k_1=2.2$ the cycle is more regular while for 1.5 it is quite 
erratic. We show the toroidal field (i.e.\ the component $B_y$ in 
the direction
of the mean shear flow) at an arbitrarily chosen mesh point as well as
its squared value (which could be taken as a proxy of the
sunspot number), the mean magnetic energy in the full domain, as well
as its contributions from the mean and fluctuating fields.
In all cases, the magnetic field is normalized by the equipartition
value, $\Beq=\sqrt{\mu_0\rho_0}\urms$, where $\rho_0=\bra{\rho}$ is the
mean density, which is conserved for periodic and shearing--periodic boundary
conditions.
Here, angular brackets denote volume averages and overbars
are defined as $xy$ averages, so
\EQ
\meanBB(z,t)=\int\BB(x,y,z,t)\;\dd x\;\dd y/L_x L_y,
\EN
which implies that
\EQ
\bra{\BB^2}=\bra{\meanBB^2}+\bra{\bb^2}.
\EN
One sees that $\bra{\bb^2}$ shows fluctuations that are not strongly
correlated with the variations of the mean field (\Fig{pncomp2}).
This is important because the lack of a correlation is sometimes used 
to argue that the Sun's small-scale magnetic field must be created by a
local small-scale dynamo and disconnected with the large-scale dynamo.

\begin{figure}[t!]\begin{center}
\includegraphics[width=\columnwidth]{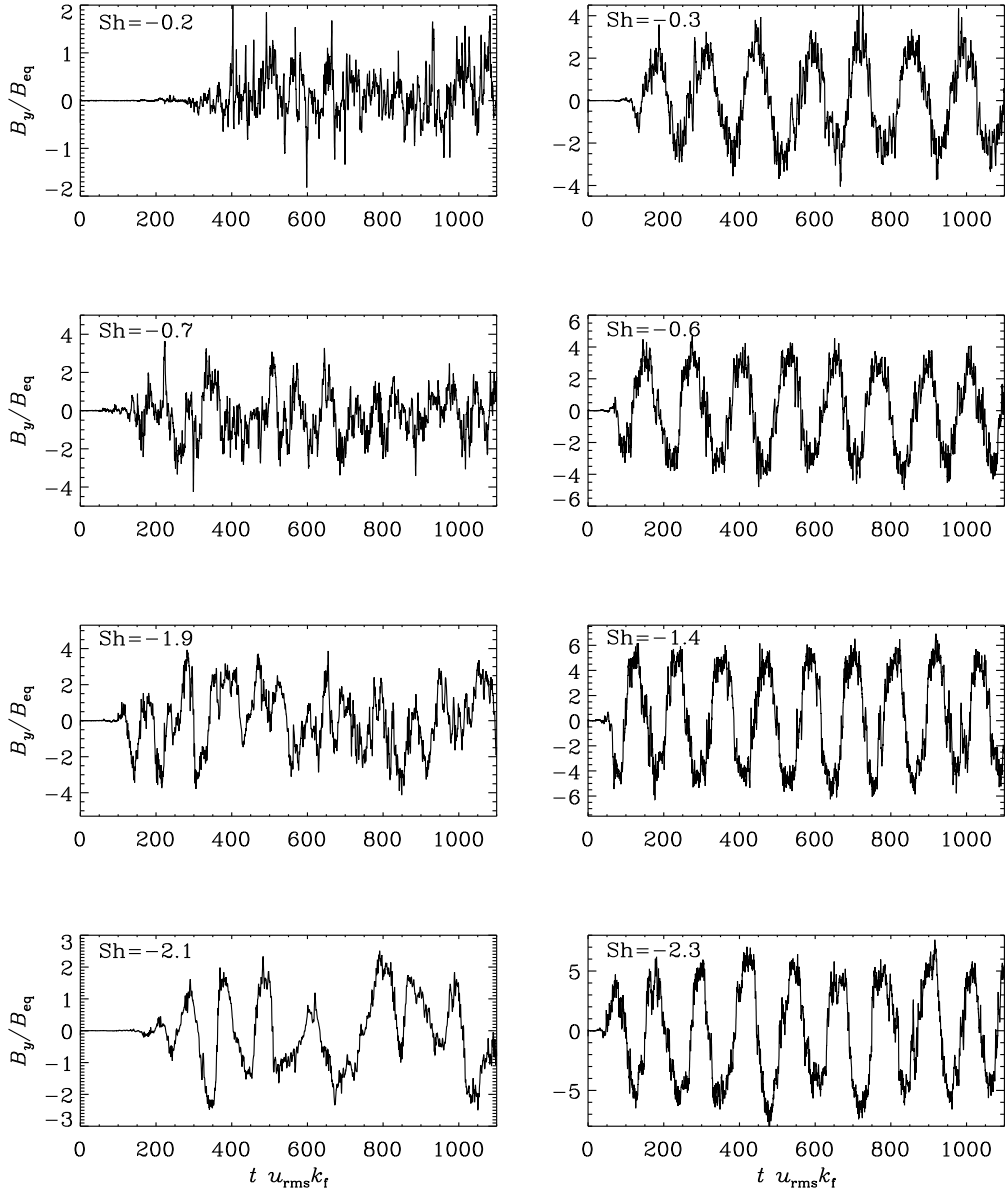}
\end{center}\caption[]{
Time series of $B_y/\Beq$ for $\kf/k_1=1.5$ (left column) and
$\kf/k_1=2.2$ (right column) for different values of $\Sh$.
Again, $\Rm=22$ for $\kf/k_1=1.5$ (left column) and 
$\Rm=9$ for $\kf/k_1=2.2$ (right column),and $\Pm=5$ in both cases.
For $\kf/k_1=2.2$ the oscillations tend to become less sinusoidal
for larger values of $\Sh$, while for $\kf/k_1=1.5$ there are
strong fluctuations that tend to become somewhat weaker for larger
values of $\Sh$.
}\label{pncompb}\end{figure}

There are several other interesting differences between the two cases.
The cycle period is given by $\omega_{\rm cyc}\approx\etat k_1^2$
\citep{KB09}, and with $\etat\approx\urms/3\kf$ \citep{SBS08} we have
$\omega_{\rm cyc}\approx\onethird\urms\kf(k_1/\kf)^2$; thus the normalized
cycle period is $T_{\rm cyc}\urms\kf\approx2\pi\urms\kf/\omega_{\rm cyc}
\approx6\pi(\kf/k_1)^2\approx91$ for $\kf/k_1=2.2$, which agrees with
the result shown in the upper right panel of \Fig{pncomp2}.
Next, for $\kf/k_1=2.2$ the mean magnetic energy and $\bra{\BB^2}$ are
about 4 times larger than for $\kf/k_1=1.5$.
This value is larger than the one expected from the theory where
this ratio should be
equal to the ratio of the respective values of $\kf$ \citep{BB02}, namely
$2.2/1.5\approx1.5$.

Next, we ask how the properties of the dynamo change as it becomes
more supercritical.
This is shown in \Fig{pncompb} where we plot time series of $B_y/\Beq$
for $\kf/k_1=1.5$ and $\kf/k_1=2.2$ for different values of $\Sh$.
Note that for $\kf/k_1=2.2$ the oscillations become less sinusoidal
for larger values of $\Sh$, while for $\kf/k_1=1.5$ there are strong
fluctuations that become somewhat weaker for larger values of $\Sh$.

\begin{figure}[t!]\begin{center}
\includegraphics[width=\columnwidth]{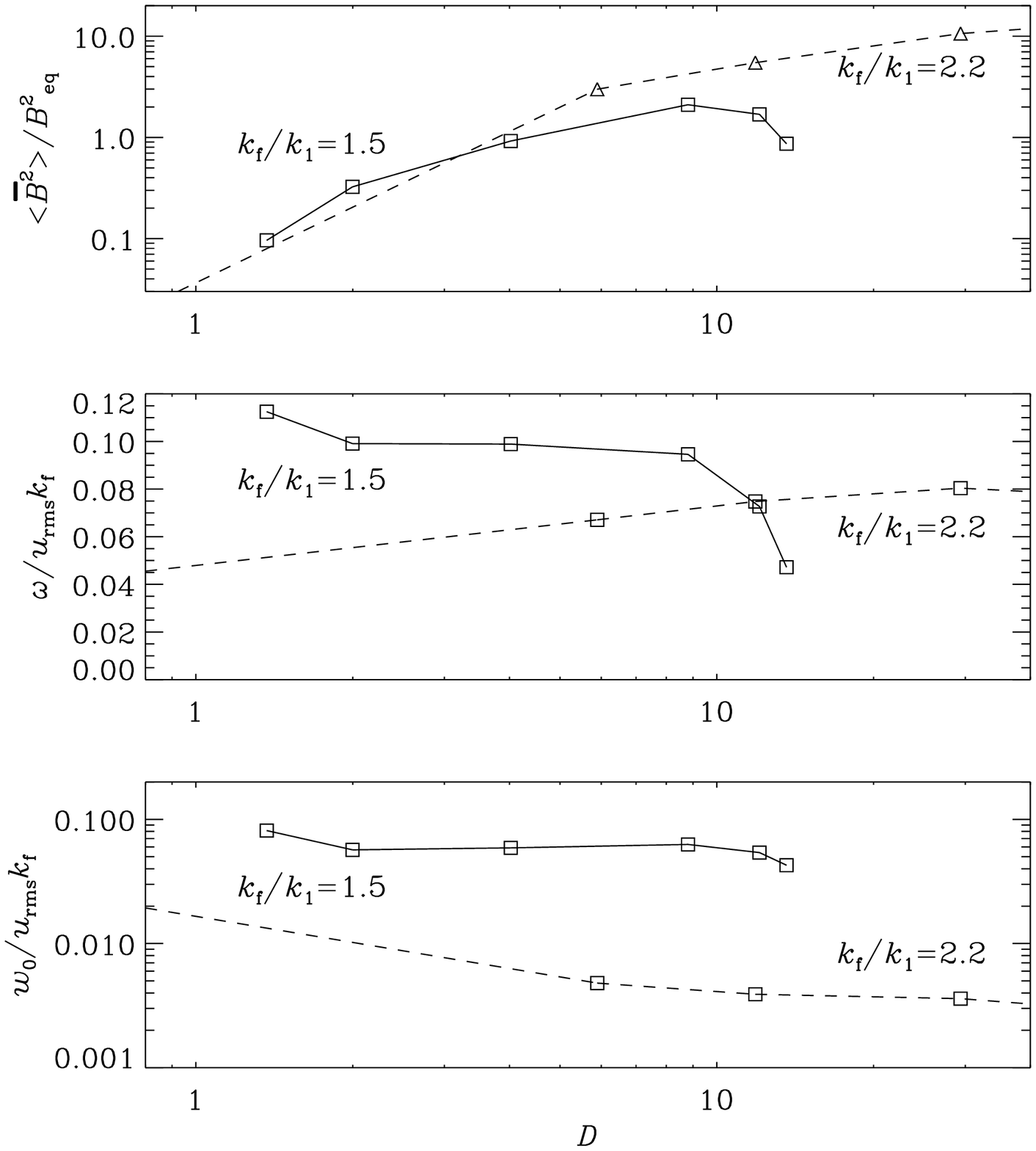}
\end{center}\caption[]{
Scaling of the relative magnetic energy of the mean field,
$\bra{\meanBB^2}/\Beq^2$, normalized cycle frequency,
and normalized quality versus dynamo $D=C_\alpha C_S$.
}\label{ppen}\end{figure}

A more quantitative way of assessing the properties of the
large-scale dynamo is by looking at the scaling of the magnetic
energy of the mean field and the cycle frequency as a function of the
nominal dynamo number, $D=C_\alpha C_S$, where
\EQ
C_\alpha=\alpha/\etaT k_1=\iota\epsf\kf/k_1,\quad
C_S=S/\etaT k_1=3\iota\,\Sh\,(\kf/k_1)^2,\quad
\EN
are non-dimensional numbers measuring the expected value of the
$\alpha$ effect (assuming $\alpha\approx\onethird\tau\bra{\oo\cdot\uu}$,
$\etat\approx\onethird\tau\bra{\uu^2}$, with $\tau=(\urms\kf)^{-1}$
and $\bra{\oo\cdot\uu}\approx\kf\bra{\uu^2}$), as well as the shear 
or $\Omega$ effect.
In these approximations, $\etaT=\eta+\etat$ is the expected total magnetic
diffusivity and $\iota=(1+3/\Rm)^{-1}$ is a correction factor that takes
into account finite conductivity effects resulting from the fact that
$\etat\neq\etaT$.  
The results are shown in \Fig{ppen}.

For $\kf/k_1=1.5$, the scaling of $\bra{\meanBB^2}/\Beq^2$ with $D$
suggests that the critical value is between 1 and 2, i.e., somewhat smaller
than the theoretical value of 2 \citep{BS05}.
For $\kf/k_1=2.2$ the critical value is $<1$.
The cycle frequencies are approximately independent of $D$, except that
for $\kf/k_1=1.5$ there is a sharp drop for $D>10$.
Owing to fluctuations, a Fourier spectrum of the time series is not sharp
but has a certain width.
We determine the quality or width, $w_0$, by fitting the spectrum to a
Gaussian proportional to $P(\omega)\sim\exp[-(\omega-\omega_0)^2/2 w_0^2]$.
Also the values of $w_0$, shown in the last panel of \Fig{ppen}, are
approximately independent of $D$.
For $\kf/k_1=2.2$, $w_0$ is substantially smaller than for $\kf/k_1=1.5$.
This indicates that the cycle period is better defined for larger
scale separation ratios.

\section{Active regions and their inclination angle}
\label{sunspots}

One should expect that the sunspot number depends in a complicated
way on the magnetic field strength.
If sunspots are indeed relatively shallow phenomena, the field must
be locally concentrated to field strengths of up to $3\kG$.
A candidate for a mechanism that can concentrate mean fields of $\sim300\G$,
which is about 10\% of the local equipartition field strength, is the
negative effective magnetic pressure instability (NEMPI).
This is a remarkable phenomenon resulting from the suppression of
turbulent pressure by a moderately strong large-scale magnetic field.
This suppression is stronger than the added magnetic pressure from the
mean field itself, so the net effect is a negative one.

The fact that this phenomenon can lead to an instability in a stratified
layer was first found in mean-field models \citep{BKR10,BKKR11,KBKMR11},
and more recently in DNS \citep{BKKMR11}.
However, NEMPI has not yet been able to explain
flux concentration in the direction along the mean magnetic field, i.e.,
the large-scale structures remain essentially axisymmetric.
To discuss the theoretical origin of this, we need to look at the
underlying mean-field theory.
Similar to the effective magnetic diffusivity in the mean electromotive
force, the sum of Reynolds and Maxwell stresses from the small-scale field
depends on the mean magnetic field in a way that looks like a Maxwell stress
from the mean field, but with renormalized coefficients.
The concept of expressing the Reynolds stress from the fluctuating
velocities, $\overline{u_i u_j}$, by the mean flow $\meanUU$ is of
course familiar and leads to the usual turbulent viscosity term,
$-\nut(\meanU_{i,j}+\meanU_{j,i})$.
However, in the presence of a mean magnetic field, symmetry arguments
allow one to write down additional components, in particular those
proportional to $\delta_{ij}\meanBB^2$ and $\meanB_i\meanB_j$.
The sum of Reynolds and Maxwell stresses from the fluctuating velocity
and magnetic fields is given by
\EQ
\overline{\Pi}^{\rm f}_{ij}\equiv\meanrho\,\overline{u_i u_j}
-\overline{b_i b_j}/\mu_0+\half\overline{\bb^2}/\mu_0,
\EN
where the superscript f indicates contributions from the fluctuating field.
Expressing $\overline{\Pi}^{\rm f}_{ij}$ in terms of the mean field, the
leading terms are \citep{KRR90,KR94,KMR96,RK07}
\EQ
\overline{\Pi}^{\rm f}_{ij}=\qs\meanB_i\meanB_j/\mu_0
-\half\qp\delta_{ij}\meanBB^2/\mu_0+...
\EN
where the dots indicate the presence of additional terms that enter
when the effects of stratification affect the anisotropy of the turbulence
further.
Note in particular the definition of the signs of the terms involving
the functions $\qs(\meanBB)$ and $\qp(\meanBB)$.
This becomes obvious when writing down the mean Maxwell stress resulting
from both mean and fluctuating fields, i.e.,
\EQ
-\meanB_i\meanB_j/\mu_0+\half\delta_{ij}\meanBB^2/\mu_0
+\overline{\Pi}^{\rm f}_{ij}
=-(1-\qs)\meanB_i\meanB_j/\mu_0+\half(1-\qp)\delta_{ij}\meanBB^2/\mu_0+...
\EN
A broad range of different DNS have now confirmed that $\qp$ is positive
for $\Rm>1$, but $\qs$ is small and negative.
A positive value of $\qs$ (but with large error bars) was originally
reported for unstratified turbulence \citep{BKR10}.
Later, stratified simulations with isothermal stable stratification
\citep{BKKR11} and convectively unstable stratification \citep{KBKMR11}
show that it is small and negative.
Nevertheless, $\qp(\meanBB)$ is consistently positive provided $\Rm>1$
and $\meanB/\Beq$ is below a certain critical value that is around 0.5.
This implies that it is probably not possible to produce flux
concentrations stronger than half the equipartition field strength.
So, making sunspots with this mechanism alone is maybe unlikely.

The significance of a positive $\qs$ value comes from mean-field
simulations with $\qs>0$ indicating the formation of three-dimensional
(non-axisymmetric) flux concentrations \citep{BKR10}. This result was 
later identified to be a direct consequence of having $\qs>0$ \citep{KBKR11}.
Before making any further conclusions, it is important to assess the
effect of other terms that have been neglected.
Two of them are related to the vertical stratification, i.e.\ terms
proportional to $g_ig_j$ and $g_i\meanB_j+g_j\meanB_i$ with $\grav$
being gravity. The coefficient of the former
term seems to be small \citep{KBKMR11} and the second only has an effect
when there is a vertical imposed field.
However, there could be other terms such as $\meanJ_i\meanJ_j$ as well as
$\meanJ_i\meanB_j$ and $\meanJ_j\meanB_i$ that have not yet been looked at.

Yet another alternative for causing flux concentrations is the suppression
of turbulent (convective) heat transport which might even be strong
enough to explain the formation of sunspots \citep{KM00}.
It is conceivable that effects form heat transport become more important
near the surface, so that a combination of the negative effective magnetic
pressure and the suppression of turbulent heat transport are needed.
Another advantage of the latter is that this mechanism works for vertical
fields and is isotropic with respect to the horizontal plane, so one should
expect the formation of three-dimensional non-axisymmetric structures.

\begin{figure}[t!]\begin{center}
\includegraphics[width=\columnwidth]{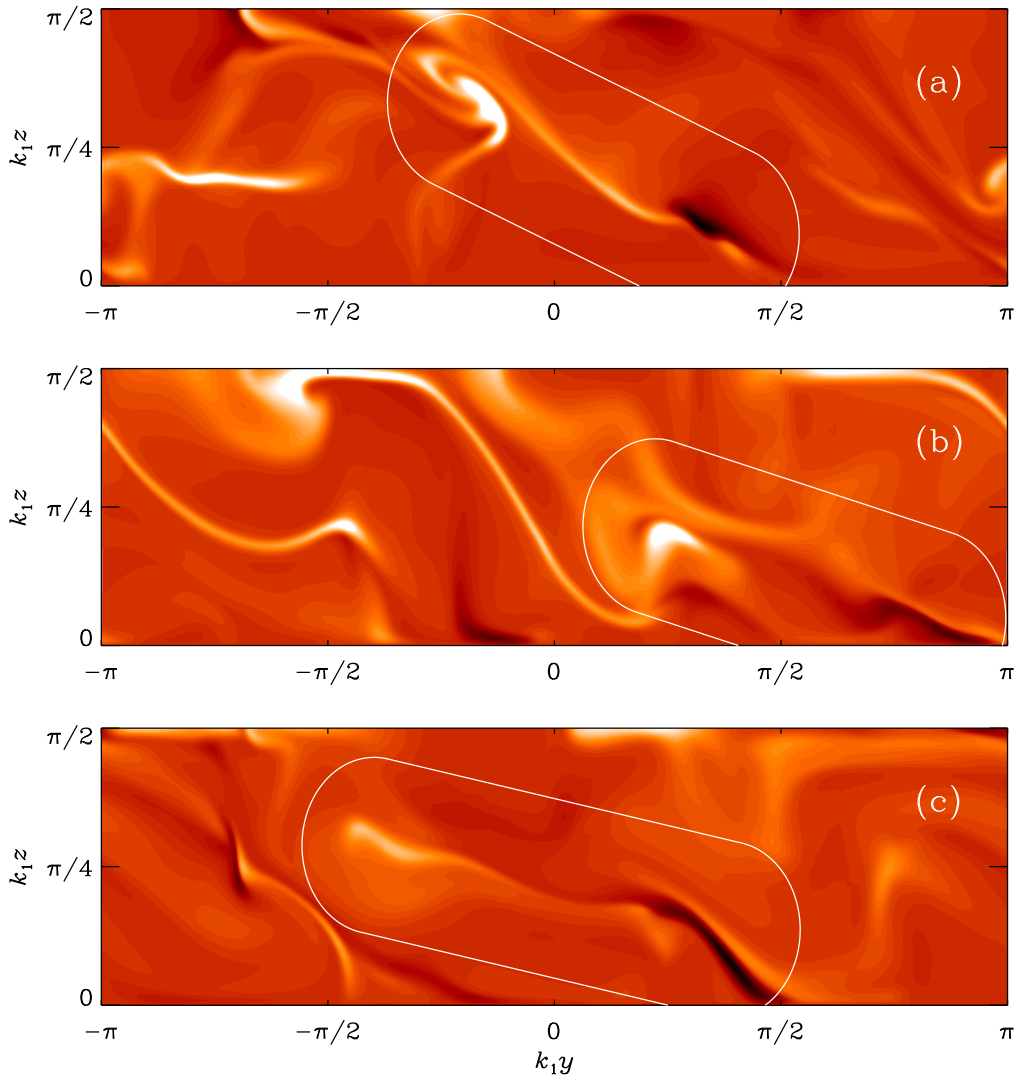}
\end{center}\caption[]{
Magnetograms of the radial field at the outer surface
on the northern hemisphere at different times for a simulation
presented in \cite{B05}.
Light shades correspond to field vectors pointing out of the domain,
and dark shades correspond to vectors pointing into the domain.
The elongated rings highlight the positions of bipolar regions.
Note the clockwise tilt relative to the $y$ (or toroidal) direction,
and the systematic sequence of polarities (white left and dark right)
corresponding to $\meanB_y>0$.
Here, the $z$ direction corresponds to latitude.
}\label{pmagnetogram}\end{figure}

Once an active region forms and it is bipolar,
we must ask ourselves how to explain the observed tilt angle.
This question cannot be answered within the framework of NEMPI
alone, but it requires a connection with the underlying dynamo.
Here, we can refer to the work of \cite{B05} where bipolar regions
occur occasionally at the surface of a domain in which shear-driven
turbulent dynamo action was found to operate; see \Fig{pmagnetogram}.
The reason for the tilt is here not the Coriolis force, as is usually
assumed, but shear; see also \cite{KS08}.
In the simulations of \cite{B05}, shear was admittedly rather strong
compared with the turbulent velocity, so the effect is exaggerated compared
to what we should expect to happen in the Sun.
However, even then there are a few other problems.
One of them is that the bipolar regions appear usually quite far away
from each other (\Fig{pmagnetogram}).
This may not be realistic.
On the other hand, it is not clear how to scale this model to the Sun.
In this model, the scale separation ratio is rather small, so the extent
of the bipolar regions is comparable to a few times the convective eddy
size, which, for the solar surface, is not very big (a few Mm).
However, on that small scale one would not expect the effects of
differential shear to be very important.

\begin{figure}[t!]\begin{center}
\includegraphics[width=\columnwidth]{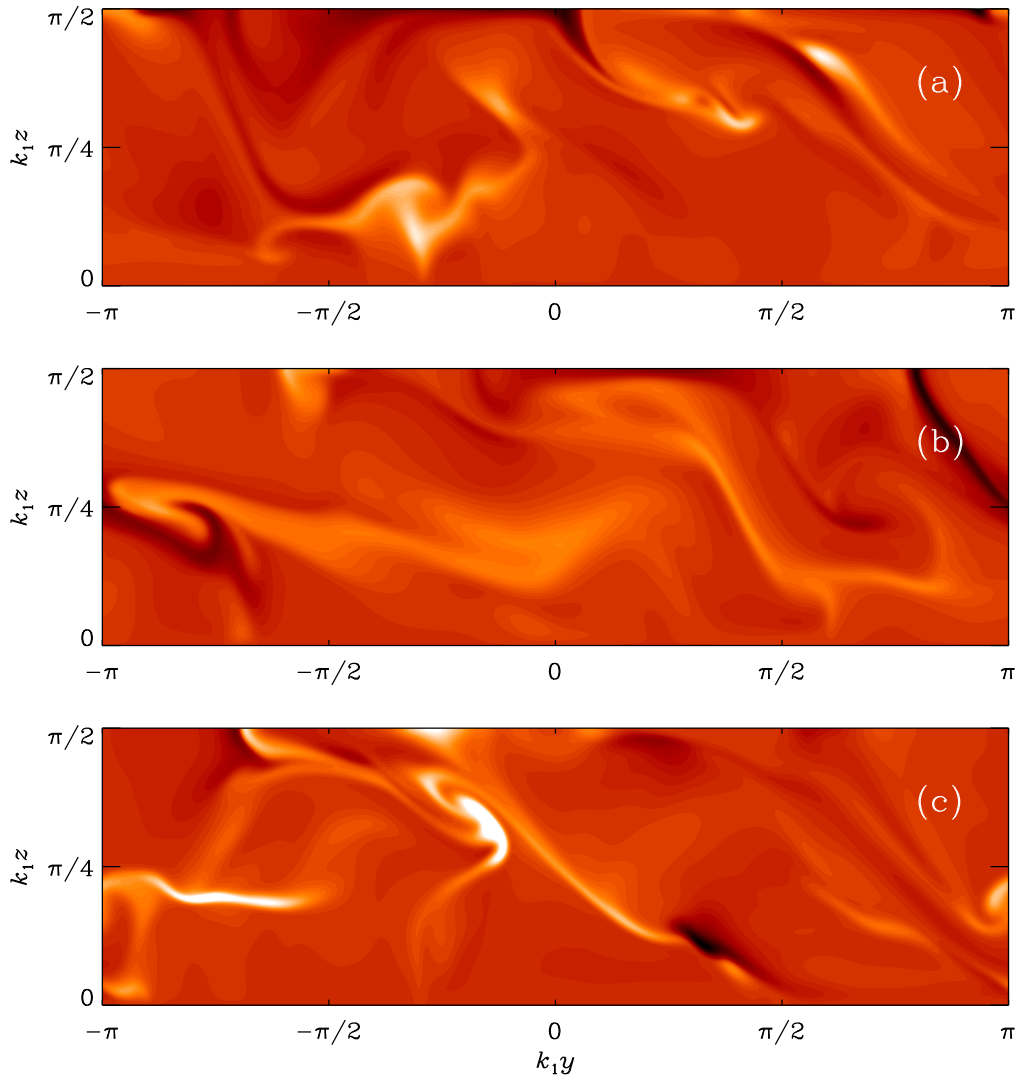}
\end{center}\caption[]{
Similar to \Fig{pmagnetogram}a, showing also the results 
4 and 2 turnover times earlier. The last panel is identical 
to \Fig{pmagnetogram}a.
}\label{pmagnetogram2}\end{figure}

Another issue is that in the model of \cite{B05}, bipolar regions occur
only occasionally.
To illustrate this, we show in \Fig{pmagnetogram2} the resulting magnetograms
for three times that are separated by about 2 turnover times.
Clearly, other structures can appear too and the field is not always bipolar.

\section{Conclusions}

It is clear that the magnetohydrodynamics of mean magnetic and velocity
fields is quite rich and full of important effects.
The standard idea that sunspots and bipolar regions form as a result
of an instability in the tachocline \citep{GD00,CDG03,PM07} may need
to be re-examined in view of several new alternative proposals being
on the horizon.
In addition to comparing model with observations at the solar surface,
there are ways of comparison both beneath the surface and above.
Particularly exciting are the recent determinations of \cite{IZK11}
of some sort of activity at $\approx60\Mm$ depth.
It is also of interest to explain magnetic activity in the solar wind,
and especially its magnetic helicity which has recently been found to
be bi-helical, i.e., of opposite signs at large and small length scales
\citep{BSBG11} and positive at small length scales in the north.
This is particularly interesting, because such a result has recently
been reproduced by distributed dynamo simulations of \cite{WBM11} who
also find positive magnetic helicity at small length scales in the north.
More detailed and varied comparisons between the different approaches
are thus required because we can more fully understand the Sun's
activity cycles and their long term variations.

\acknowledgments
We acknowledge the allocation of computing resources provided by the
Swedish National Allocations Committee at the Center for
Parallel Computers at the Royal Institute of Technology in
Stockholm and the National Supercomputer Centers in Link\"oping.
This work was supported in part by
the European Research Council under the AstroDyn Research Project
227952 and the Swedish Research Council grant 621-2007-4064.


\begin{discussion}

\discuss{Mu\~noz-Jaramillo}{
If sunspots are produced at the surface, which processes would lead to
their decay?
}

\discuss{Brandenburg}{
I think it could be the continued submersion of magnetic structures.
The system remains time-dependent and new structures will form near
the surface, while old ones disappear from view.
}

\discuss{Luhmann}{
Axel, you are one of the few dynamo modelers that include the
corona and larger heliosphere in your models and thinking.
How important is that to the results of the models, and do
models not including that aspect have compromised results? 
}

\discuss{Brandenburg}{
The magnetic helicity flux divergence is crucial for alleviating
catastrophic alpha quenching; see the next talk by Candaleresi
et al. (2011). Helicity fluxes through surface carry about 30%
of the total;  the rest goes through the equator.
The observed magnetic helicity spectra support our understanding
in terms of the magnetic helicity evolution equation.
Models not including helicity fluxes suffer artificially strong
catastrophic quenching, but only if their magnetic Reynolds numbers
are really large.
}

\discuss{Choudhuri}{
Flux rise simulations based on the idea that the toroidal field
forms in the tachocline explained Joy's law other characteristics
of sunspot graphs. Can these results be recovered if sunspots form
from near-surface fields?
}

\discuss{Brandenburg}{
The tilt angles of near-surface produced flux concentrations
is determined by latitudinal shear, as was demonstrated by
Brandenburg (2005).
}

\discuss{Nandi}{
For the near-surface, shallow dynamo to work, you need
processes that are slow enough to store, amplify, and
transport fields on 11 year timescales.
However, in the upper convection zone, eddy turnover timescales
are short.
Can you comment on how you reconcile this with your
shallow dynamo?
}

\discuss{Brandenburg}{
Magnetic helicity conservation is generally responsible for
prolonging the time scales.
In fact, the partial alleviation of catastrophic quenching
by magnetic helicity fluxes means that the timescales are
not infinite.
The dynamo period is proportional to
$(\alpha\partial\Omega/\partial r)^{-1/2}$,
so $\alpha$ quenching prolongs the period.
}

\end{discussion}

\end{document}